%% file: Paper.tex
\newcommand{\CLASSINPUTtoptextmargin}{0.75in}%
\newcommand{\CLASSINPUTbottomtextmargin}{1in}%
\newcommand{\CLASSINPUTinnersidemargin}{0.63in}%
\newcommand{\CLASSINPUToutersidemargin}{0.63in}%
\documentclass[conference,10pt,letterpaper]{IEEEtran}%
\usepackage{amsmath}
\usepackage{graphicx}
\usepackage{siunitx}
\usepackage{multirow}
\usepackage[none]{hyphenat}
\usepackage{float}
\usepackage{subfig}
\usepackage{dblfloatfix}
\usepackage{tikz}
\usepackage{textcomp}
\usepackage{hyperref}
\usepackage{lipsum}

\newcommand\copyrighttext{%
  \footnotesize \textcopyright 2012 IEEE. Personal use of this material is permitted.
  Permission from IEEE must be obtained for all other uses, in any current or future
  media, including reprinting/republishing this material for advertising or promotional
  purposes, creating new collective works, for resale or redistribution to servers or
  lists, or reuse of any copyrighted component of this work in other works.
  }
\newcommand\copyrightnotice{%
\begin{tikzpicture}[remember picture,overlay]
\node[anchor=south,yshift=10pt] at (current page.south) {\fbox{\parbox{\dimexpr\textwidth-\fboxsep-\fboxrule\relax}{\copyrighttext}}};
\end{tikzpicture}%
}
\usepackage{t1enc}
\usepackage{times}
\input{IMS_modify_IEEEtran_18b_CTAN_V6}

\begin{document}
\raggedbottom
%
%
%
\title{Tight Non-Radiating Bends of 3D-Printed Dielectric Image Lines  \\ Based on Electromagnetic Bandgap Mirrors}
%
%
%
\IMSthispaperforfinalpublication
\IMSauthor{%
\IMSauthorblockNAME{
Leonhard Hahn\IMSauthorrefmark{\#1},
Tobias Bader\IMSauthorrefmark{\#},
Christian Carlowitz\IMSauthorrefmark{\#},
Martin Vossiek\IMSauthorrefmark{\#} and 
Gerald Gold\IMSauthorrefmark{\#}
}
\\%
\IMSauthorblockAFFIL{
\IMSauthorrefmark{\#}Institute of Microwaves and Photonics (LHFT), Friedrich-Alexander-Universität Erlangen-Nürnberg
}
\\%
\IMSauthorblockEMAIL{
\IMSauthorrefmark{1}leonhard.hahn@fau.de
}
}
%
\maketitle
\copyrightnotice
%
%
%
\begin{abstract}
This paper reports on a novel compact, low-loss bending technique for additively manufactured dielectric image lines between 140\,GHz and 220\,GHz. 
Conventional bending approaches require either large curvature radii or technologically challenging permittivity variations to prevent radiation loss at line discontinuities, e.g. caused by narrow bends. In contrast, this work uses extremely compact, easy-to-manufacture electromagnetic bandgap (EBG) cells to solve the afore mentioned challenge for the first time. These offer excellent reflection properties and thus enable broadband and low-loss (IL~<~1\,dB) guidance of electromagnetic waves by means of total reflection. Without increasing the complexity of the process, both the high-pass behaviour and the enormous space requirement of conventional dielectric bends are completely avoided. In addition, the use of EBGs improves the mutual isolation of dielectric image lines of up to 30\,dB. Therefore, a promising solution for the realization of narrow, 3D-printed, low-loss signal distribution networks in the sub-THz domain is offered.

\end{abstract}
%
%

\section{Introduction}
The sub-THz range is a promising region for future radar systems due to its unique characteristics. Compared to the millimeter wave (mmW) range, its small wavelength allows larger bandwidths, which significantly improves the resolution of imaging radar systems. However, unlike optical systems, it retains a robust measurement method that allows it to penetrate materials or operate in bad weather. For this reason, THz FMCW radar systems are focus of research \cite{Starke.2022,Kissinger.2021}. To further increase the potential of the THz domain, radar systems with multi-channel signal acquisition are used. Since line attenuation increases significantly in the THz range, the interconnection of several spatially distributed receivers requires transmission lines with minimal losses. Dielectric image lines (DILs) are an excellent option for this purpose, as their planar nature also provides improved mechanical robustness and polarisation stability compared to conventional dielectric waveguides \cite{Tesmer.2022,King.1958}.
In comparison to metallic waveguides, DILs are also significantly lighter and smaller in size \cite{Tesmer.2020}. Additionally, additive manufacturing allows for extremely low cost and easy production and eliminates the need for lossy gluing \cite{Tesmer.2021}. However, one limitation when using DILs is their tendency to exhibit parasitic radiation at discontinuities~\cite{Hahn.2024}. Thus, both low-loss and small dielectric networks cannot be realized if large bending angles are needed, or if only very limited space is available, as with chip-to-chip interconnects. Previous attempts to overcome this problem have included the use of large bending radii \cite{Hahn.2024}, high permittivity dielectrics close to the discontinuity \cite{Elio.1998}, or the selective deceleration of phase fronts inside the DIL using permittivity gradients \cite{Shiina.1986}. However, those conventional approaches have drawbacks:
\begin{figure}[t]
\centering
\includegraphics[width=80mm]{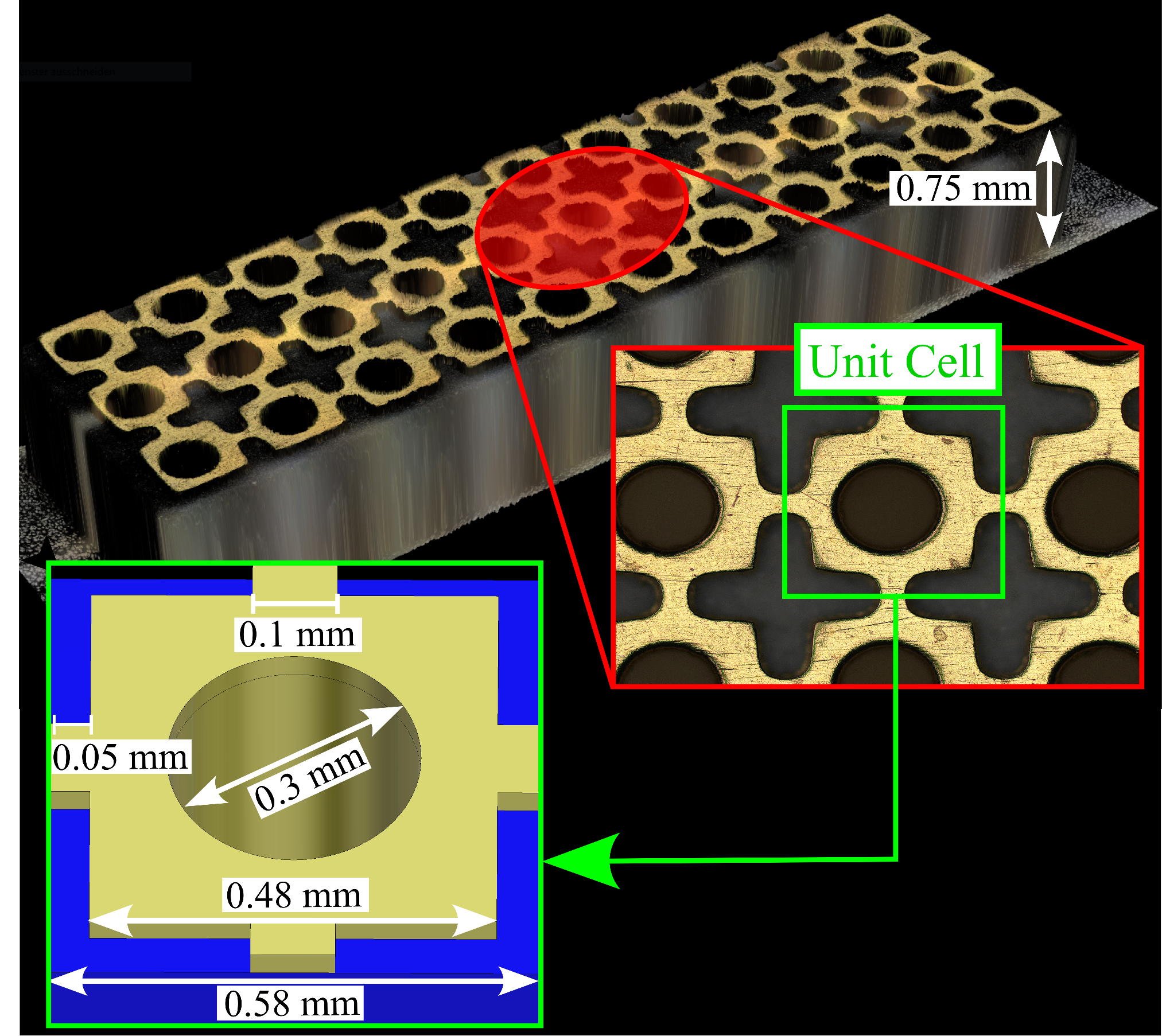}
\caption{Microscopic 3D-image of the EBG used with surface close-up and dimensions of its corresponding unit cell.}
\label{fig:EBG_Overall}
\end{figure}
Large radii reduce radiation but increase cable length, which in turn increases losses. Phase deceleration \cite{Shiina.1986} is used in optics, but its application in the sub-THz range is challenging because the bending losses exhibit high-pass behaviour and become more dominant at lower frequencies. In addition, this implementation increases complexity. Another approach is the use of total reflection, where electromagnetic waves are guided through narrow radii with low losses \cite{Ogusu.1985}. This concept is dependent on highly reflective surfaces; one approach could be the use of purely metallic mirrors.\\
However, a novel approach is presented in this paper: Electromagnetic bandgap materials (EGB) have proven to be a suitable reflector for planar mmW antennas \cite{Lau.2011}. When compared to metal, they offer similar reflection magnitudes with zero-phase reflection at lower mmW frequencies \cite{Dey.2022}. This work now utilizes this approach and tailors ideal reflector EBG structures for the first time to the sub-THz range. Low-loss bends for DILs without restrictions on bending radii or space requirements were designed, manufactured and characterized.


\section{EBG Unit Cell}
\begin{figure}[t!]
\centering
\includegraphics[width=80mm]{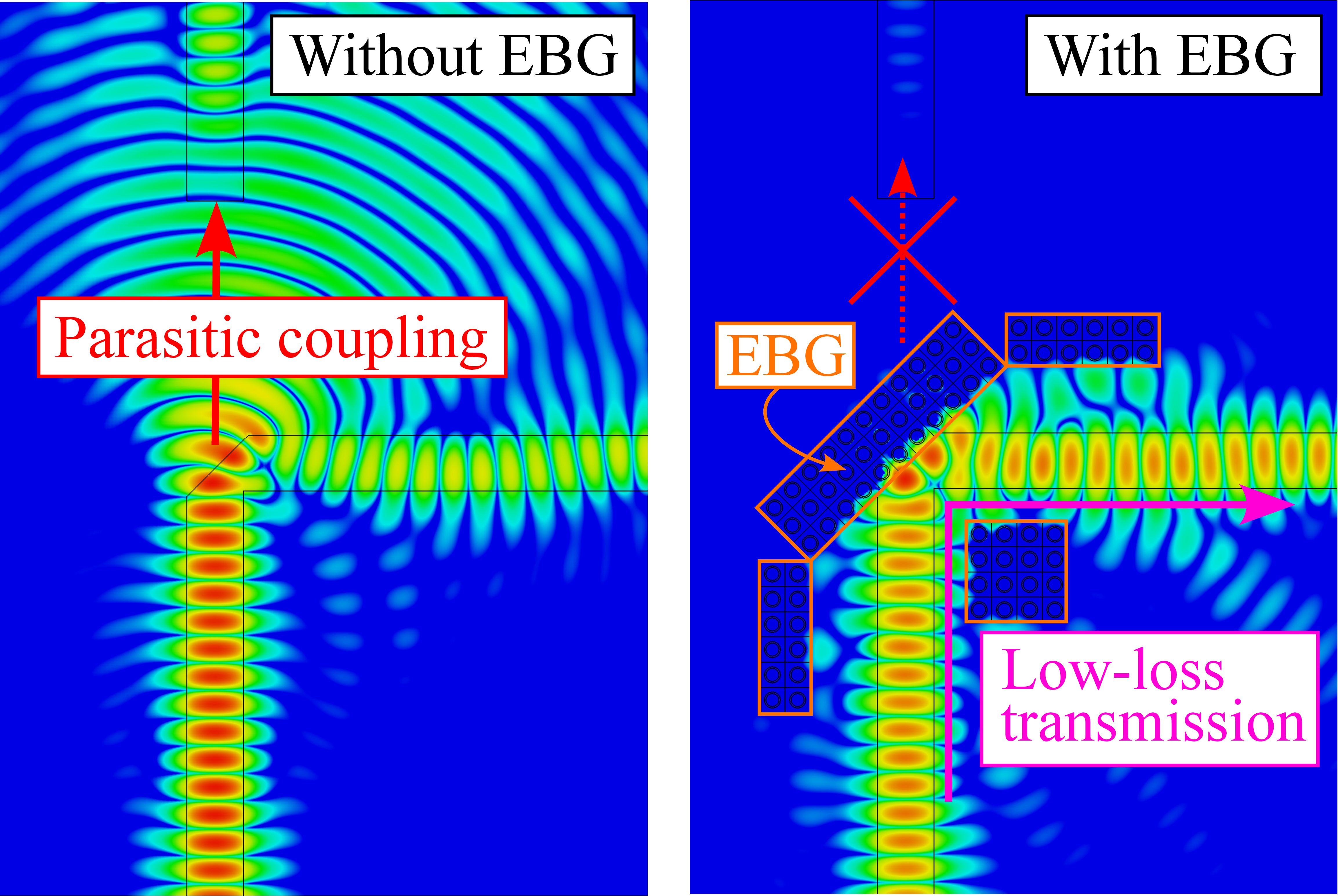}
\caption{3D field simulation of the EBG for estimation of insertion loss and isolation capability for an incident $\mathrm{HE_{11}}$ fundamental mode.}
\label{fig:Felder_Sim}
\end{figure}
EBGs are commonly understood as a metallo-dielectric structure, consisting of a unit cell which is periodically repeated to form a lattice. One single cell is much smaller than the wavelength, which enables a description with lumped elements. By structuring the top metal, a high surface impedance can be achieved, preventing the propagation of electromagnetic waves in certain frequency bands. This work is based on mushroom EBGs, which consist of square top metal patches connected to the bottom metal by vias, see Fig.~\ref{fig:EBG_Overall}. These structures have proven to be particularly suitable for TE and TM surface wave suppression and are comparatively easy to fabricate \cite{Kovacs.2012}. The patches and their resulting current loops form a serial inductance; their distance from neighbouring patches defines a serial capacitance, resulting in filtering behaviour. In contrast to conventional mushroom EBGs, simulative optimizations were performed to improve the reflection factor between \SI{140}{\giga\hertz} and \SI{220}{\giga\hertz}. Therefore, connections between the patches are used to further increase serial inductance. The final EBG is shown in Fig.~\ref{fig:EBG_Overall}.
The EBG's dielectric core consists of Rogers RO3003 with a thickness of \SI{0.75}{\milli\meter}, relative permittivity $\varepsilon_{\text{r}}=3.00$ and loss tangent $\tan\delta=10^{-3}$. For low-loss reflection, the centre of the dielectric line must be exposed, since this is the location of the electric field maximum of its $\mathrm{HE_{11}}$ mode, as illustrated by the inset in Fig.~\ref{fig:MeasSetup}.
This is achieved by chamfering the DIL. In order to realize a reflection around a \SI{90}{\degree} corner, the EBG mirror must be tilted by \SI{45}{\degree} so that the centres of the adjacent DILs coincide at a common point on the mirror. By inserting an EBG within this chamfer, an incoming wave is reflected, since the $\mathrm{HE_{11}}$ mode cannot propagate through the EBG. This also benefits the isolation capability of these lines, as adjacent DILs are sensitive to parasitic radiation, especially towards radiation from the same angle of incidence. The described setup is verified by electromagnetic field simulations. The simulated field distribution in Fig.~\ref{fig:Felder_Sim} confirms that an incoming $\mathrm{HE_{11}}$ fundamental mode wave is almost completely reflected with the help of the EBG structures, so that low-loss transmission around the corner and high isolation from the neighbouring line can be achieved. The main EBG (3x11 unit cells) within the DIL's chamfer is sufficient for the concept's functionality, for further investigation of surface wave propagation, 3 additional side elements (2x6 / 4x4 unit cells) are used.


\section{Measurement Setup}\label{sec:MeasSetup}
\begin{figure}[t!]
\centering
\includegraphics[width=80mm]{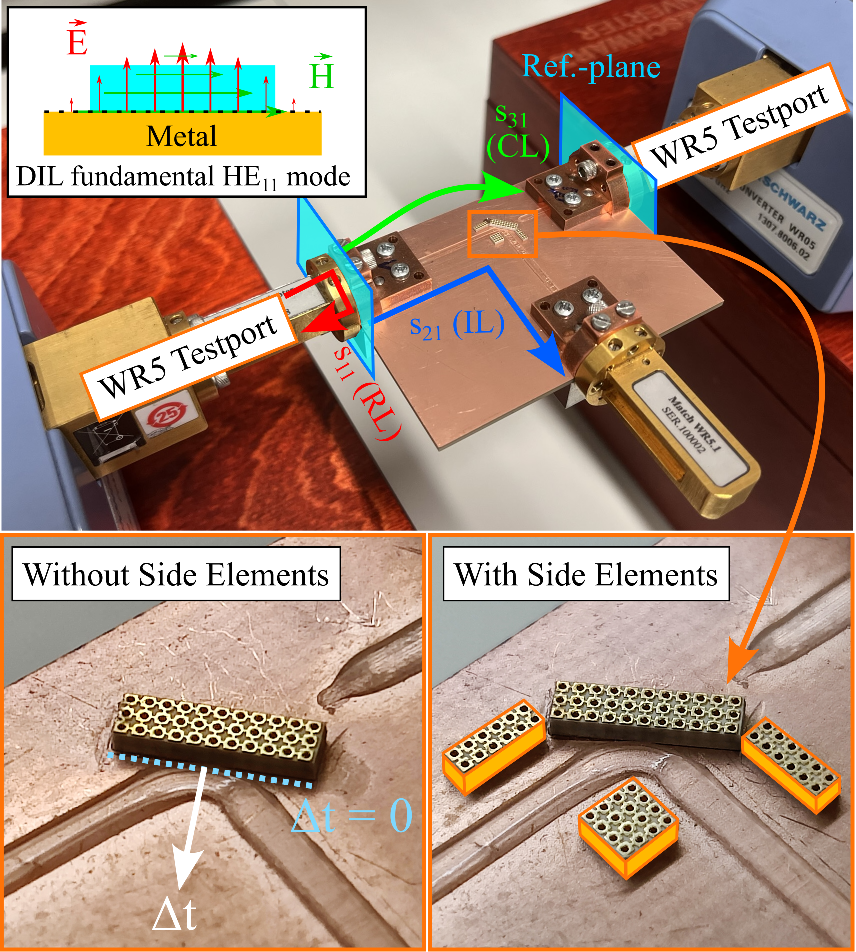}
\caption{Measurement setup for the characterization of return loss ($s_{11}$), insertion loss ($s_{21}$) and coupling loss ($s_{31}$) of different DIL bends. The top left inset visualizes the DIL operation principle including $\text{HE}_{11}$ mode. $\Delta \text{t}$ denotes possible variations of chamfer depth departing from the DIL center.}
\label{fig:MeasSetup}
\end{figure}
To determine the effect of EBGs on dielectric bends, a \SI{90}{\degree} DIL bend was fabricated using FDM-based 3D-printing and it was characterized before and after the insertion of EBG elements within its chamfer. In addition, measurements with and without side elements are carried out, to further quantify their impact, see Fig.~\ref{fig:MeasSetup}. For reference, various other bending techniques are printed and analyzed, such as chamfered and rounded DILs with variable bending radii. The measurement setup including the DIL's fundamental mode $\text{HE}_\text{11}$ can be seen in Fig.~\ref{fig:MeasSetup}, consisting of two Rohde\&Schwarz ZVA220 frequency converters for S-parameter measurements from \SI{140}{\giga\hertz} to \SI{220}{\giga\hertz}. Low-loss mode converters from \cite{Hahn.2024} are used to feed the DILs with the WR5 test port adapters. After TOSM calibration, the waveguide flanges are used as reference plane for subsequent measurements, see Fig.~\ref{fig:MeasSetup}.
The geometric dimensions of the DILs with a width of $a \approx \SI{1.3}{\milli\meter}$ and height $b \approx \SI{0.3}{\milli\meter}$ correspond to the previously described parameters from \cite{Hahn.2024}. The matching of the bends $s_{11}$ is determined at the input port, $s_{21}$ describes the combined losses due to line length (dielectric loss, ohmic loss) and bending radius (radiation loss). DIL intereference is determined by $s_{31}$ with the help of an additional sniffer line, which quantifies parasitic radiation in case of a worst-case scenario of an adjacent and aligned DIL. Thereby, the feasibility of compact, low-loss DIL bends using reflective EBG elements can be investigated.


\section{Results}
\begin{figure}[t]
\centering
\includegraphics[width=85mm]{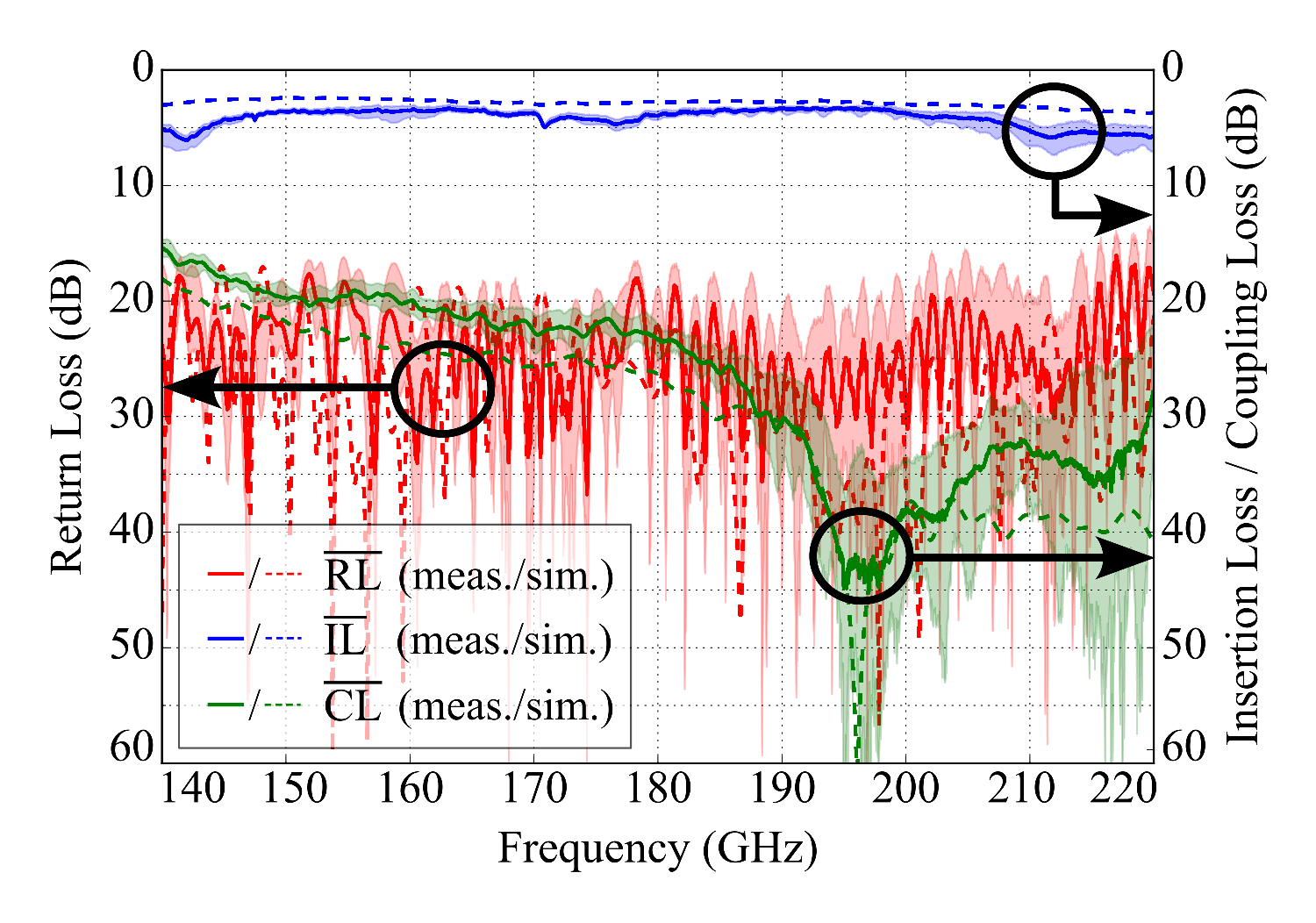}
\caption{Simulated and measured S-parameters of \SI{90}{\degree} DIL bend with EBGs, including an investigation of measurement variation for 10 different EBG samples.}
\label{fig:Materialschwankung}
\end{figure}
\begin{figure}[b!]
\centering
\includegraphics[width=85mm]{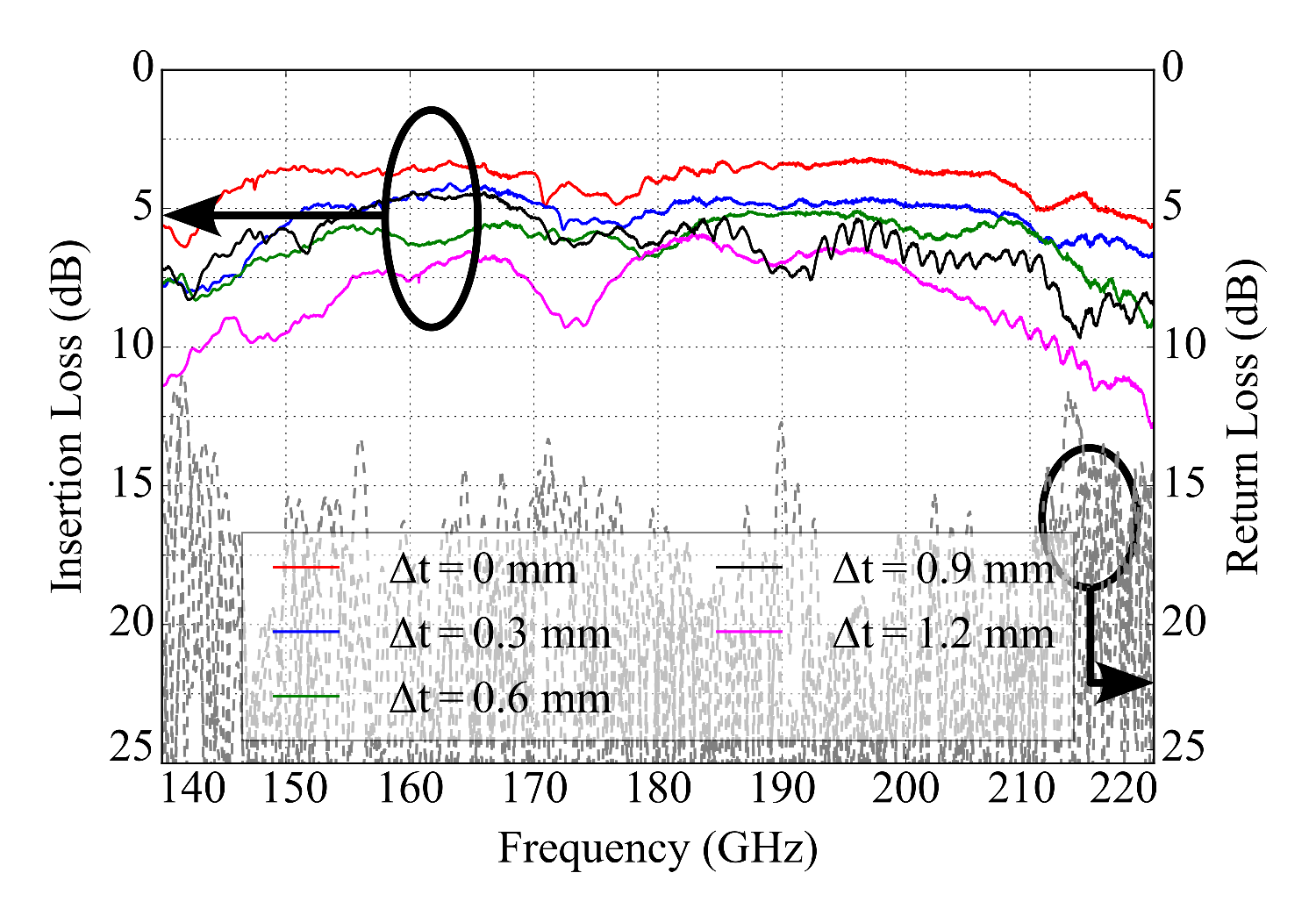}
\caption{Influence of different chamfers depths on measured insertion loss and coupling loss, including a comparison to bends without use of EBGs.}
\label{fig:Phasentiefe}
\end{figure}
\begin{figure}[b!]
\centering
\includegraphics[width=85mm]{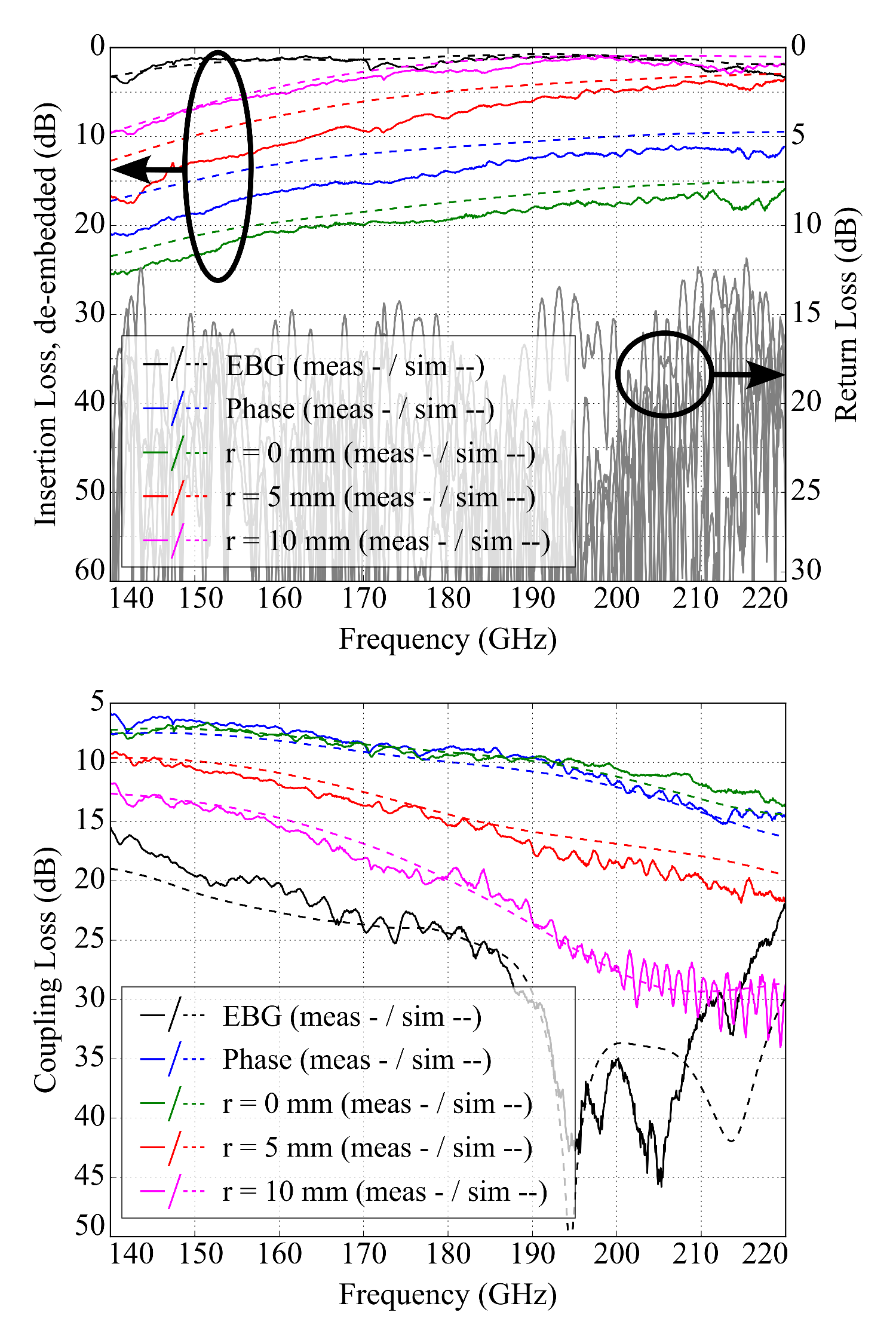}
\caption{Simulated and measured S-parameters for technology comparison of different DIL bending techniques.}
\label{fig:Technologievergleich}
\end{figure}

To describe the transmission characteristics of the EBG bends, S-parameter measurements are performed according to the measurement setup from Sec.~\ref{sec:MeasSetup}, see Fig.~\ref{fig:Materialschwankung} for the simulation and measurement results. For evaluation of the influence of manufacturing errors, the same DIL bend was measured with 10 different EBG samples. Fig.~\ref{fig:Materialschwankung} shows the mean value of the magnitude of $s_{11}$, $s_{21}$ and $s_{31}$ over all 10 samples, the colored areas indicate its maximum spread. It can be observed that all DIL bends are matched over the entire frequency band. The return loss (RL) varies between \SI{20}{\decibel} and \SI{30}{\decibel} for the major part of the spectrum and reaches \SI{16}{\decibel} at worst. The insertion loss (IL) is approximately constant up to \SI{200}{\giga\hertz} at approximately \SI{3.4}{\decibel}, with additional line and mode-conversion losses being included. The coupling loss (CL) shows that parasitic coupling is reduced by at least \SI{15}{\decibel}, between \SI{190}{\giga\hertz} and \SI{220}{\giga\hertz} the isolation even averages between \SI{30}{\decibel} and \SI{45}{\decibel}. Especially for IL, the achieved spread is small and measurements are in very good agreement with the simulations. 
\begin{figure}[t!]
\centering
\includegraphics[width=85mm]{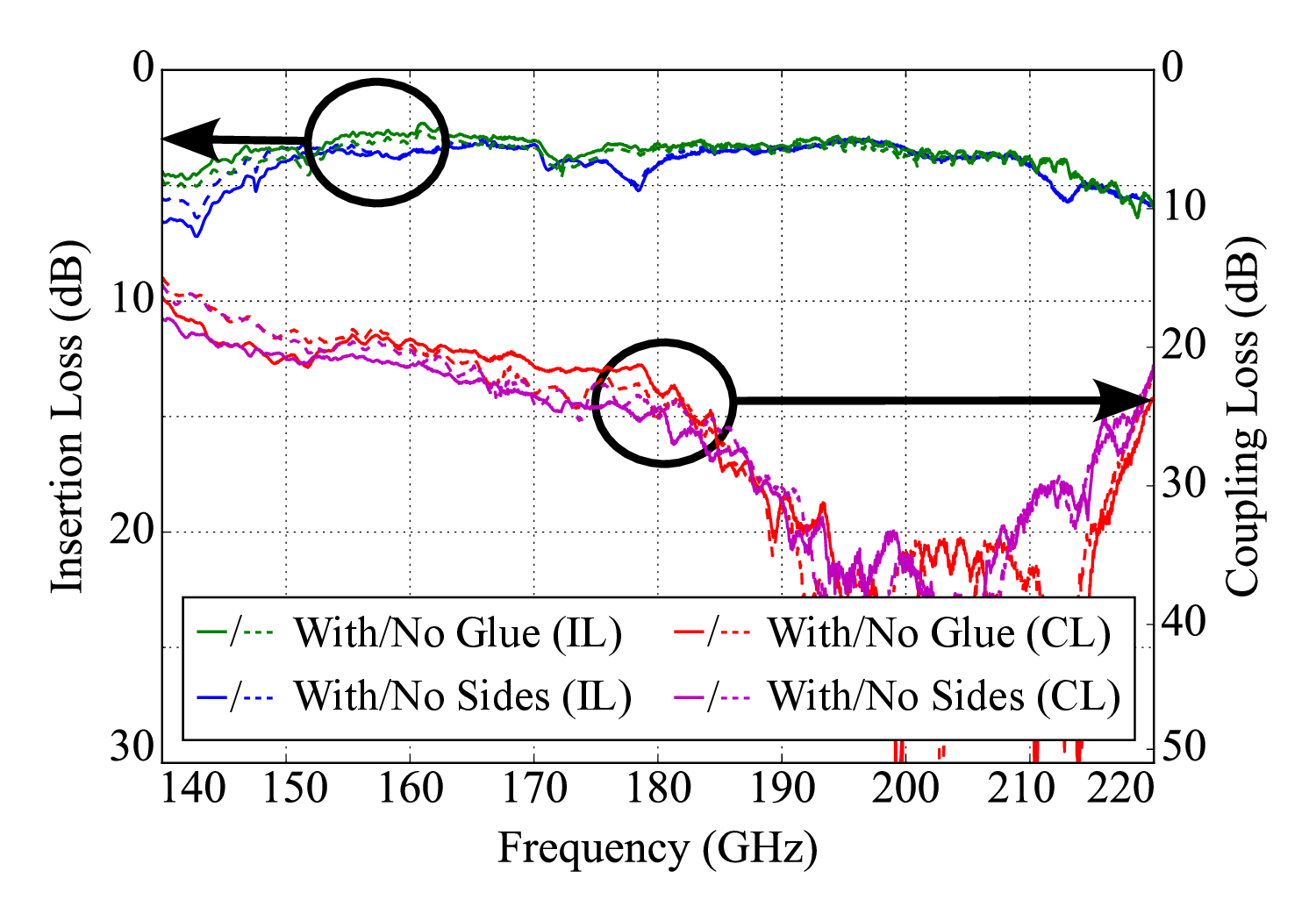}
\caption{Impact of gluing and side elements on measured insertion loss and coupling loss.}
\label{fig:Gluing}
\end{figure}
In addition, the impact of side EBGs and the adhesive, which is required for EBG attachment, is shown in Fig.~\ref{fig:Gluing}. It is evident that neither IL nor CL undergo significant changes with the insertion of side EBGs or the adhesive bond. The EBG reflector's position is determined by the depth of the chamfer $\Delta t$, thus, $\Delta t$ is varied and its influence on the loss is shown in Fig.~\ref{fig:Phasentiefe}. If $\Delta t = 0$, the chamfer connects the center of both lines at one common point, whereas $\Delta t > 0$ cuts deeper, as seen in Fig.~\ref{fig:MeasSetup}. Chamfering depths with $\Delta t \neq 0$ were expected to increase IL as an incident wave is not reflected into the centre of the \SI{90}{\degree} adjacent DIL. It is also noticeable that an EBG placed on the corner of a small radius instead of a chamfer yields comparably good results, as long as the EBG is inserted close to the centre of both approaching DILs. Regardless of the selected chamfer depth, a satisfactory matching $\text{RL}>\SI{15}{\decibel}$ is maintained.
Finally, a comparison between different bending techniques is conducted. Five different options are compared, including three bending radii, where $r=\SI{0}{\milli\meter}$ refers to a very compact, almost rectangular bend, while $r>\SI{0}{\milli\meter}$ refers to a streched curve with less discontinuity. In addition, a chamfered DIL bend and, finally, the proposed EBG bend are examined in the study. Fig.~\ref{fig:Technologievergleich} illustrates both the simulated and measured RL, IL and CL. The insertion loss (IL) is de-embedded and describes only the additional loss due to radiation from the bend, as the properties of line length and mode converters noted in \cite{Hahn.2024} have been subtracted. All bends are matched independently, with a minimum return loss (RL) of \SI{12}{\decibel}. The losses diminish by over \SI{15}{\decibel} towards larger bending radii ($r=\SI{0}{\milli\meter} \rightarrow r=\SI{10}{\milli\meter}$). However, they do require more space. Chamfered bends do not require any extra space, but they reduce the insertion loss by approximately \SI{5}{\decibel} only, with a high remaining attenuation of \SI{12}{\decibel} at \SI{200}{\giga\hertz}.
In contrast, an EBG bend also requires minimal space, but significantly reduces losses over the entire frequency band by \SI{15}{\decibel} to over \SI{20}{\decibel}, resulting in an extremely low $\text{IL}<\SI{1}{\decibel}$ at \SI{200}{\giga\hertz}. A radius $r = \SI{10}{\milli\meter}$ achieves no IL improvements; for ranges between \SI{140}{\giga\hertz} and \SI{170}{\giga\hertz} even larger radii are required to achieve similarly low attenuation. There is also a significant improvement in coupling loss with the use of EBGs. Predictably, compact bends possess strong coupling due to their sudden discontinuities and strong radiation, this behaviour is gradually mitigated as the radius increases. In contrast, EBG bends achieve maximum isolation, especially between \SI{190}{\giga\hertz} and \SI{210}{\giga\hertz}. In comparison to similarly compact bending techniques, an isolation improvement between \SI{10}{\decibel} and \SI{30}{\decibel} is achieved. The measurements are in excellent agreement with the simulation results, especially for the EBG variants.


\section{Conclusion}

In this paper, dielectric bends with minimized losses and space requirements were realized with the help of EBGs. The EBG cells possess excellent reflection properties that enable low-loss and high-matching guidance of incident $\text{HE}_{11}$ fundamental modes around narrow \SI{90}{\degree}~corners. Herefore, proper EBG alignment is crucial to ensure the reflection of incoming waves into the center of adjacent DILs. Chamfers or narrow bending radii are especially suitable for achieving this requirement. The use of adhesive has no negative impact on the electrical performance. Side EBGs show no significant influence in measurements. After de-embedding, the EBG bends display a reproducibly low loss of down to \SI{0.9}{\decibel} at \SI{200}{\giga\hertz}. In contrast to conventional bending radii, the space requirements remain minimal. Furthermore, the isolation properties of the DILs are maximized by the use of EBGs. Improved transmission properties are only possible by using significantly bigger radii, which, however, conflicts with the desired compactness. In future work, this approach is reusable for the realization of complex, yet compact components (splitters, couplers) for large DIL based distribution networks.


\section*{Acknowledgment}


\newcommand{\IMSacktext}{%
This work was supported by the German Research Foundation (DFG) within the frame of the research project TeraCaT (funding code 1519-54792 7558172). We would also like to thank Tim~Pfahler for the inspiring discussions.
}

\IMSdisplayacksection{\IMSacktext}


\bibliographystyle{IEEEtran}
\bibliography{bibfile}{}

\end{document}

%% file: IMS_modify_IEEEtran_18b_CTAN_V6.tex
\makeatletter

\def\@IMSauthorblockNAMEstyle{\normalfont\IMSauthorsize}
\def\@IMSauthorblockAFFILstyle{\normalfont\IMSaffilsize}
\def\@IMSauthorblockEMAILstyle{\normalfont\IMSaffilsize}
\def\IMSauthorblockNAME#1{%
\relax\@IMSauthorblockNAMEstyle%
#1%
}%
\def\IMSauthorblockAFFIL#1{%
\relax\@IMSauthorblockAFFILstyle%
\vskip\@IEEEauthorblockAtopspace
#1%
}%
\def\IMSauthorblockEMAIL#1{%
\relax\@IMSauthorblockEMAILstyle%
\vskip\@IEEEauthorblockAtopspace
#1%
}%
\newcommand{\IMSauthor}[1]{%
\ifIsBlindReviewVersion%
\author{\phantom{\parbox{\textwidth}{\center\relax#1}}}%
\else%
\author{\parbox{\textwidth}{\center\relax#1}}%
\fi%
}%
\newif\ifIsBlindReviewVersion

\def\IMSthispaperforfinalpublication{\IsBlindReviewVersionfalse}
\def\@maketitle{\newpage
\bgroup\par\addvspace{0.5\baselineskip}\centering%
\ifCLASSOPTIONtechnote
   {\bfseries\large\@IEEEcompsoconly{\sffamily}\@title\par}\vskip 1.3em{\lineskip .5em\@IEEEcompsoconly{\sffamily}\@author
   \@IEEEspecialpapernotice\par{\@IEEEcompsoconly{\vskip 1.5em\relax
   \@IEEEtitleabstractindextextbox{\@IEEEtitleabstractindextext}\par
   \hfill\@IEEEcompsocdiamondline\hfill\hbox{}\par}}}\relax
\else
   \vskip0.2em{\IMStitlesize\ifCLASSOPTIONtransmag\bfseries\LARGE\fi\@IEEEcompsoconly{\sffamily}\@IEEEcompsocconfonly{\normalfont\normalsize\vskip 2\@IEEEnormalsizeunitybaselineskip
   \bfseries\Large}\@title\par}\vskip1.0em\par
   \ifCLASSOPTIONconference%
      {\@IEEEspecialpapernotice\mbox{}\vskip\@IEEEauthorblockconfadjspace%
       \mbox{}\hfill\begin{@IEEEauthorhalign}\@author\end{@IEEEauthorhalign}\hfill\mbox{}\par}\relax
   \else
      \ifCLASSOPTIONpeerreviewca
         {\@IEEEcompsoconly{\sffamily}\@IEEEspecialpapernotice\mbox{}\vskip\@IEEEauthorblockconfadjspace%
          \mbox{}\hfill\begin{@IEEEauthorhalign}\@author\end{@IEEEauthorhalign}\hfill\mbox{}\par
          {\@IEEEcompsoconly{\vskip 1.5em\relax
           \@IEEEtitleabstractindextextbox{\@IEEEtitleabstractindextext}\par\hfill
           \@IEEEcompsocdiamondline\hfill\hbox{}\par}}}\relax
      \else
         \ifCLASSOPTIONtransmag
           {\@IEEEspecialpapernotice\mbox{}\vskip\@IEEEauthorblockconfadjspace%
            \mbox{}\hfill\begin{@IEEEauthorhalign}\@author\end{@IEEEauthorhalign}\hfill\mbox{}\par
           {\vspace{0.5\baselineskip}\relax\@IEEEtitleabstractindextextbox{\@IEEEtitleabstractindextext}\vspace{-1\baselineskip}\par}}\relax
         \else
           {\lineskip.5em\@IEEEcompsoconly{\sffamily}\sublargesize\@author\@IEEEspecialpapernotice\par
           {\@IEEEcompsoconly{\vskip 1.5em\relax
            \@IEEEtitleabstractindextextbox{\@IEEEtitleabstractindextext}\par\hfill
            \@IEEEcompsocdiamondline\hfill\hbox{}\par}}}\relax
         \fi
      \fi
   \fi
\fi\par\addvspace{0.0\baselineskip}\egroup}

\def\IMStitlesize{\@setfontsize{\IMStitlesize}{18}{21pt}}
\def\IMSauthorsize{\@setfontsize{\IMSauthorsize}{12}{13pt}}
\def\IMSaffilsize{\@setfontsize{\IMSaffilsize}{12}{13pt}}
\def\IMScaptionsize{\@setfontsize{\IMScaptionsize}{8}{9pt}}
\def\IMSbibsize{\@setfontsize{\IMSbibsize}{8}{9pt}}

\def\@IEEEauthorblockNstyle{\IMSauthorsize\@IEEEcompsocnotconfonly{\sffamily}\@IEEEcompsocconfonly{\large}}
\def\@IEEEauthorblockAstyle{\IMSaffilsize\@IEEEcompsocnotconfonly{\sffamily}\@IEEEcompsocconfonly{\itshape}\@IEEEcompsocconfonly{\large}}
\def\@IEEEauthordefaulttextstyle{\IMSauthorsize\@IEEEcompsocnotconfonly{\sffamily}\sublargesize}

\def\thebibliography#1{\section*{\refname}%
    \addcontentsline{toc}{section}{\refname}%
    \IMSbibsize\@IEEEcompsocconfonly{\small}\vskip 0.3\baselineskip plus 0.1\baselineskip minus 0.1\baselineskip
    \list{\@biblabel{\@arabic\c@enumiv}}%
    {\settowidth\labelwidth{\@biblabel{#1}}%
    \leftmargin\labelwidth
    \advance\leftmargin\labelsep\relax
    \itemsep \IEEEbibitemsep\relax
    \usecounter{enumiv}%
    \let\p@enumiv\@empty
    \renewcommand\theenumiv{\@arabic\c@enumiv}}%
    \let\@IEEElatexbibitem\bibitem%
    \def\bibitem{\@IEEEbibitemprefix\@IEEElatexbibitem}%
\def\newblock{\hskip .11em plus .33em minus .07em}%
\ifCLASSOPTIONtechnote\sloppy\clubpenalty4000\widowpenalty4000\interlinepenalty100%
\else\sloppy\clubpenalty4000\widowpenalty4000\interlinepenalty500\fi%
    \sfcode`\.=1000\relax}

\long\def\@makecaption#1#2{%
\ifx\@captype\@IEEEtablestring%
\par\@IEEEtabletopskipstrut
\else
\@IEEEfigurecaptionsepspace
\fi
\setbox\@tempboxa\hbox{\normalfont\IMScaptionsize {#1.}\nobreakspace\nobreakspace #2}%
\ifdim \wd\@tempboxa >\hsize%
\setbox\@tempboxa\hbox{\normalfont\IMScaptionsize {#1.}\nobreakspace\nobreakspace}%
\parbox[t]{\hsize}{\normalfont\IMScaptionsize\noindent\unhbox\@tempboxa#2}%
\else
\ifCLASSOPTIONconference \hbox to\hsize{\normalfont\IMScaptionsize\hfil\box\@tempboxa\hfil}%
\else \hbox to\hsize{\normalfont\IMScaptionsize\box\@tempboxa\hfil}%
\fi\fi
\ifx\@captype\@IEEEtablestring%
\@IEEEtablecaptionsepspace
\else
\fi}

\newlength\tablecaptiontotableskip
\newlength\figuretocaptionskip
\def\@IEEEfigurecaptionsepspace{\vskip\figuretocaptionskip\relax}%
\def\@IEEEtablecaptionsepspace{\vskip\tablecaptiontotableskip\relax}%

\def\abstract{\normalfont%
\@IEEEabskeysecsize\bfseries\textit{\abstractname}\,\bfseries\textit{---}\,%
\@IEEEgobbleleadPARNLSP}%

\def\IEEEkeywords{\normalfont%
\@IEEEabskeysecsize\bfseries\textit{\IEEEkeywordsname}\,\bfseries\textit{---}\,%
\@IEEEgobbleleadPARNLSP}%
\def\endIEEEkeywords{\relax\vspace{0.67ex}%
\par\if@twocolumn\else\endquotation\fi%
\normalsize\normalfont}%

\DeclareRobustCommand*{\IMSauthorrefmark}[1]{\raisebox{0pt}[0pt][0pt]{\textsuperscript{\footnotesize{#1}}}}%
\def\@IEEEauthorblockNtopspace{0ex}
\def\@IEEEauthorblockAtopspace{1mm}
\def\tablename{Table}
\def\thetable{\arabic{table}}
\def\IEEEkeywordsname{Keywords}
\def\subsubsection{\@startsection{subsubsection}{3}{\z@}{1.5ex plus 1.5ex minus 0.5ex}%
{0.7ex plus .5ex minus 0ex}{\normalfont\normalsize\itshape}}%
\def\@seccntformat#1{\csname the#1dis\endcsname\relax}
\def\thesectiondis{\thesection.\hskip 0.5em}
\def\thesubsectiondis{{\hbox to\parindent{\Alph{subsection}.}}}
\def\thesubsubsectiondis{{\hbox to \parindent{\arabic{subsubsection})}}}
\def\theparagraphdis{{\hbox to \parindent{\alph{paragraph})}}}
\IEEEilabelindentA \parindent
\IEEEilabelindent \IEEEilabelindentA
\IEEEelabelindent \parindent
\IEEEdlabelindent \parindent
\IEEElabelindent \parindent

\newlength\@IMSparindent
\newcommand\IMSdisplayacksection[1]{%
\ifIsBlindReviewVersion%
\noindent\phantom{\parbox[t]{\columnwidth}{\normalbaselines\setlength{\parindent}{\@IMSparindent}{#1}\strut}}
\else%
\noindent\parbox[t]{\columnwidth}{\normalbaselines\setlength{\parindent}{\@IMSparindent}{#1}\strut}%
\fi%
}%

\makeatother